\documentstyle[prb,aps,epsf]{revtex}

\begin{document}
\draft

\title{Current noise in long diffusive SNS junctions in incoherent MAR regime.}

\author{E. V. Bezuglyi, E. N. Bratus',}
\address{B. Verkin Institute for Low Temperature Physics and
Engineering, Kharkov, 61164 Ukraine}

\author{V. S. Shumeiko, and G. Wendin}
\address{Chalmers University of Technology and G\"oteborg University,
S-41296 G\"oteborg, Sweden}

\date{\today}
\wideabs{

\maketitle

\begin{abstract}
Spectral density of current fluctuations at zero frequency is
calculated for a long diffusive SNS junction with low-resistive
interfaces. At low temperature, $T \ll \Delta$, the subgap shot noise
approaches linear voltage dependence, $S=(2/ 3R)(eV + 2\Delta)$, which
is the sum of the shot noise of the normal conductor and voltage
independent excess noise. This result can also be interpreted as the
$1/3$-suppressed Poisson noise for the effective charge $q = e(1+
2\Delta/eV)$ transferred by incoherent multiple Andreev reflections
(MAR). At higher temperatures, anomalies of the current noise develop
at the gap subharmonics, $eV = 2\Delta/n$. The crossover to the hot
electron regime from the MAR regime is analyzed in the limit of small
applied voltages.


\end{abstract}
}

During the last decade, considerable attention has been focused on the
study of current fluctuations in mesoscopic systems, especially on shot
noise which reflects the discrete nature of the electron charge and
correlations between electrons. Whereas in normal ballistic systems
with tunnel barriers the spectral density $S$ of the shot noise at zero
frequency approaches full Poissonian value $S_P = 2eI$,\cite{ballistic}
mesoscopic diffusive wires shorter than the inelastic scattering length
produce Poisson noise suppressed by a factor $1/3$. The universality of
this factor was demonstrated within different theoretical
models\cite{1/3Nagaev,1/3theory} and confirmed
experimentally.\cite{1/3exp} The inelastic scattering could change the
suppression factor,\cite{1/3Nagaev,1/3Nagaeveph} which was used in
experiments\cite{1/3inelastic} as a probe of the electron relaxation.

In hybrid normal-superconducting (NS) systems, the current transport at
subgap voltages, $eV < \Delta$, is associated with the transfer of an
elementary charge $2e$, instead of $e$, due to Andreev reflection of
quasiparticles from the NS interface, which converts two electrons
in the normal metal to a Cooper pair in the superconductor. As shown
theoretically\cite{doublingtheor} and observed in
experiments,\cite{doublingexp} such doubling of the elementary charge
leads to doubling of the subgap shot noise in NS junctions.

A more pronounced enhancement of the shot noise is expected in
superconducting tunnel and SNS junctions whose subgap
conductivity involves multiple Andreev reflections (MAR). In this case,
the effective transferred charge $q = (N_A+1) e$ increases along with
the number $N_A \sim 2\Delta/eV$ of Andreev reflections of
quasiparticle which overcomes the energy gap $2\Delta$ by elementary
steps $eV$. As a result, the shot noise spectral density $S(V)$ at low
voltages, $eV \ll \Delta$, should greatly exceed the Poissonian value
and approach a constant level, which can be estimated as $S(0) = 2qV/R$
$= 4\Delta/R$ for ballistic junctions with normal resistance $R$, and
as
\begin{equation}
\label{S(0)}
S(0) = 4\Delta / 3R
\end{equation}
for diffusive junctions taking into account the $1/3$-suppression
factor.  Alternatively, the enhanced noise may be interpreted as
``thermal'' noise generated by nonequilibrium quasiparticles within the
whole subgap region $|E| < \Delta$, which is a characteristic property
of the MAR regime.\cite{SNINS} The experimental observation of multiply
enhanced shot noise was reported for NbN-based tunnel
junctions\cite{Klapwijk} and several types of SNS junctions.\cite{Hoss}

Theoretical analysis of the shot noise in short junctions, with length
$d$ smaller than the coherence length $\xi_0= (\hbar D/\Delta)^{1/2}$
in the superconductor, was done in Refs.\ \onlinecite{QPC} and
\onlinecite{DC}. In such systems, the quantum coherence between the
electrons and retroreflected holes extends over the entire junction,
which leads to the ac Josephson effect and to non-ohmic $I$-$V$
characteristic.\cite{ac} Nevertheless, the estimate $q \sim (N_A+1) e$
for the effective transferred charge holds, although the zero-bias
spectral density of the shot noise differs from Eq.\ (\ref{S(0)}).
Along with the dc current, the shot noise reveals subharmonic gap
structure (SGS), i.e., steps in $S(V)$ or $dS/dV$ at $V = 2\Delta/ne$
($n = 1,2,\dots$). A semiclassical model of the shot noise in ballistic
point contacts was proposed in Ref.\ \onlinecite{Klapwijk}.

In this paper we analyze the current noise in long diffusive SNS
junctions, $d \gg \xi_0$. When the bias voltage is much larger than the
Thouless energy $E_{\text{Th}} = \hbar D/d^2 \ll \Delta$ ($D$ is the
diffusion coefficient), then the size of the coherent proximity regions
near the NS interfaces, $\xi_E = (\hbar D/E)^{1/2}$, is much smaller
than the junction length $d$ at all relevant energies $E \gtrsim eV$.
In this case, the Josephson effect is suppressed, and the subgap
current transport can be quantitatively described in terms of
incoherent MAR,\cite{SNINS,circuit} similar to the quasiclassical
theory for ballistic systems (OTBK\cite{OTBK}). The shot noise in the
incoherent MAR regime was analyzed in Ref.\ \onlinecite{SNINS} for a
diffusive SNS junction with a tunnel barrier inside the normal metal,
assuming the barrier to dominate the junction resistance. Under these
conditions, the shot noise is generated by tunneling electrons and can
therefore be calculated within the tunnel approach.\cite{LO} If the
resistance of the normal metal exceeds the resistance of possible
tunnel barriers within the junction, shot noise emerges due to impurity
scattering. In this case, it can be calculated within a Langevin
approach,\cite{Langevin} if we neglect the contribution of the small
proximity regions in the vicinity of the interfaces.  Following
Ref.\ \onlinecite{NS}, in which the Langevin equation was applied
to the current fluctuations in a diffusive NS junction, we derive an
expression for the current noise spectral density in SNS junctions at
zero frequency in terms of the nonequilibrium population numbers
$n^{e,h}(E,x)$ of electrons and holes within the normal metal, $0<x<d$,
\begin{equation} \label{noise_eh}
S = {2\over R} \int_0^d {dx \over d}\int_{-\infty}^\infty dE
\left[n^e\left(1-n^e\right) + n^h\left(1-n^h\right)\right].
\end{equation}
The electric current through the junction is given by
\begin{equation} \label{I1}
I ={d\over 2eR} \int_{-\infty}^\infty dE\; \partial_x (n^e - n^h).
\end{equation}

The population numbers obey the diffusion equation,
\begin{equation} \label{diff}
D{\partial^2 n \over \partial x^2} = I_{\varepsilon}(n),
\end{equation}
with the inelastic collision term $I_{\varepsilon}$. At $|E| > \Delta$,
the bo\-un\-da\-ry populations $n_{0,d}(E) =
n(E,x)\left|_{x=0,d}\right.$ are local-equi\-lib\-rium Fermi functions,
$n^{e,h}_0(E)\! =\! n_F(E)$, $n^{e,h}_d(E)\! =\! n_F(E\pm eV)$ (we use
the potential of the left electrode as the energy reference level). At
subgap energies, $|E| < \Delta$, the boundary conditions should be
modified in accordance with the mechanics of complete Andreev
reflection which equalizes the electron and hole population numbers at
a given electrochemical potential and blocks the net probability
current through the NS interface,\cite{circuit}
\begin{eqnarray}\label{boundary}
&\displaystyle n^e_0(E) = n^h_0(E), \; n^e_d(E-eV) = n^h_d(E+eV),
\nonumber \\
&\displaystyle n^{e\prime}_0(E) + n^{h\prime}_0(E)=0, \;
n^{e\prime}_d(E-eV) + n^{h\prime}_d(E+eV) = 0,
\end{eqnarray}
where $n^{\prime}_{0,d}$ are the boundary values of the electron and hole probability flows $\partial n / \partial x$.

\begin{figure}
\epsfxsize=8.5cm\epsffile{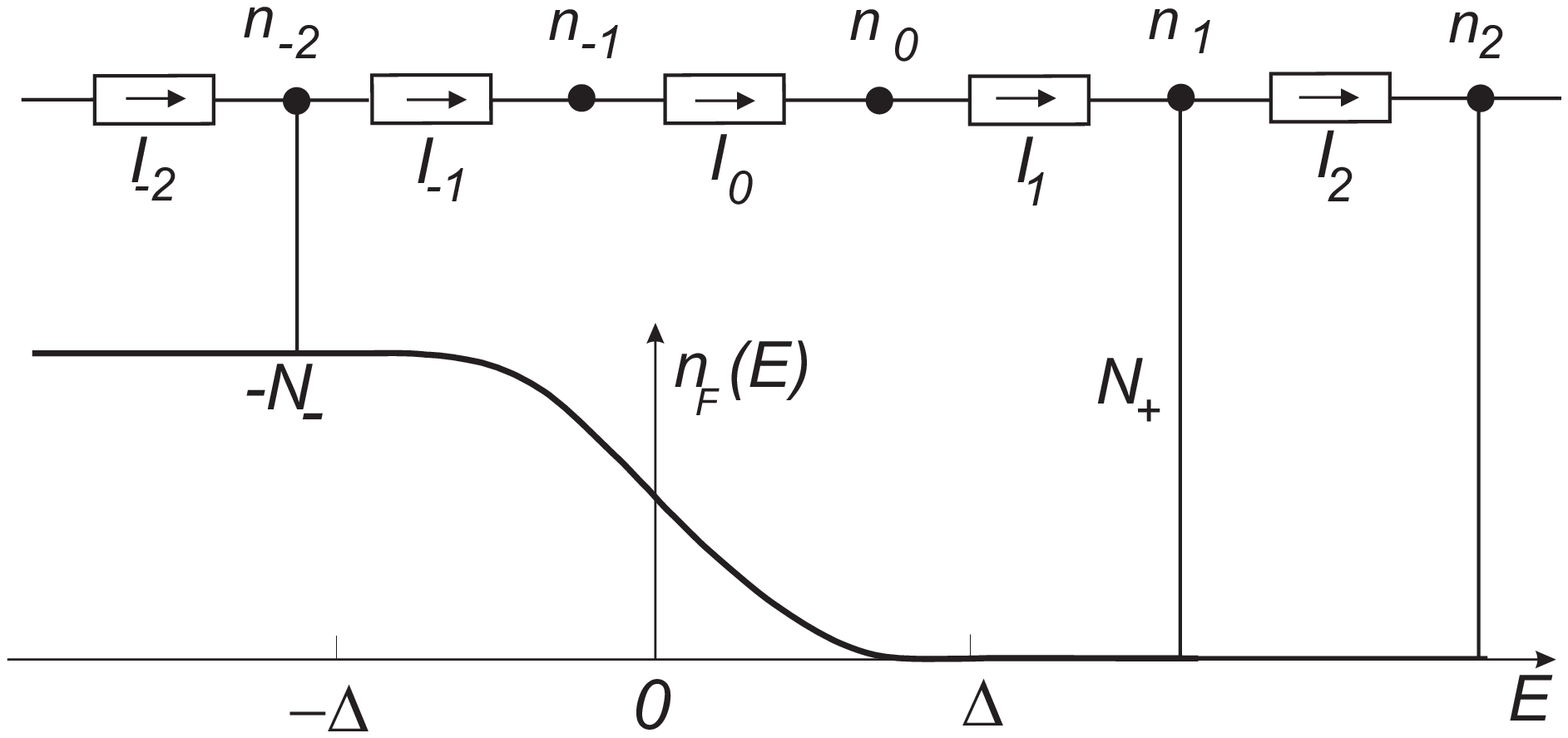} \vspace{0.25cm}
\caption{Equivalent MAR network in energy space in the limit of negligibly
small normal reflection and proximity regions at the NS interfaces.}
\label{network}
\end{figure}

In the absence of inelastic scattering, the population numbers are
linear functions of $x$, $n^{e,h}(E,x) = n^{e,h}_0(E) + x
n^{e,h\prime}(E)$, which results in the recurrences for boundary
populations and diffusive flows within the subgap region,
\begin{eqnarray} \label{recurr}
&n^{e,h}_0(E-eV) - n^{e,h}_0(E+eV) = \mp 2d n^{e,h\prime}(E\mp eV),
\\ \nonumber
&n^{e,h\prime}_0(E-eV) = n^{e,h\prime}_0(E+eV).
\end{eqnarray}
According to Ref.\ \onlinecite{circuit}, these recurrences are
equivalent to the problem of ``current'' and ``voltage'' distribution
in an equivalent network in energy space. In the present approximation,
which assumes the contribution of the proximity effect and the normal
scattering at the interfaces to be negligibly small, this network
consists of a series of resistances of unit value\cite{footnote2}
connected periodically, at the energies $E_k = E + keV$, with the
distributed ``voltage source'' $n_F(E)$ (see Fig.\ \ref{network}). The
``potentials'' $n_k$ of the network nodes with even numbers $k$
represent equal electron and hole populations $n^{e,h}_0(E+keV)$ at the
left NS interface, whereas the potentials of the odd nodes describe
equal boundary populations $n^{e,h}_d(E+keV\mp eV)$ at the right
interface. The ``currents'' $I_k$ entering $k$-th node are related to
the probability currents $n^{\prime}(E_k)$ as $I_k(E) =
-dn^{e\prime}(E_{k-1})$ (odd $k$) and $I_k(E) = dn^{h\prime}(E_{k})$
(even $k$), and represent partial electric currents transferred by the
electrons and holes across the junction, obeying Ohm's law in energy
space, $I_k = n_{k-1} - n_k$.  Within the gap, $|E_k| < \Delta$, i.e.,
at $-N_- < k < N_+$, $N_\pm(E) = \mbox{Int}[(\Delta\mp E)/eV]+1$
[$\mbox{Int}(x)$ denoting integer part of $x$], the nodes are
disconnected from the reservoir due to complete Andreev reflection and
therefore all currents flowing through the subgap nodes are equal.

Due to periodicity of the network, the partial currents obey the
relationship $I_k(E) = I_m[E+(k-m)eV]$, and the boundary population
$n_0$ is related to the node potentials $n_k$ as $n_0(E+keV) = n_k(E)$.
This allows us to reduce the integration over energy in
Eqs.\ (\ref{noise_eh}) and (\ref{I1}) to an elementary interval
$0<E<eV$,
\begin{eqnarray}
&\displaystyle I = {1\over eR} \int_0^{eV} dE \sum_{k=-\infty}^\infty I_k,
\label{I2} \\
&\displaystyle S = {2\over R} \int_0^{eV} dE \sum_{k=-\infty}^\infty\left[ 2n_k(1-n_k) +
{1\over 3}I^2_k\right].\label{S2}
\end{eqnarray}

The ``potentials'' of the nodes outside the gap, $|E_k|> \Delta$, are
equal to local-equilibrium values of the Fermi function, $n_k(E) =
n_F(E_k)$ at $k \geq N_+$, $k \leq -N_-$. The partial currents flowing
between these nodes,
\begin{equation} \label{outgap}
I_k = n_F(E_{k-1}) -n_F(E_k), \;\; k > N_+, \; k \leq -N_-,
\end{equation}
are associated with thermally excited quasiparticles. The subgap
currents may be calculated by Ohm's law for the series of $N_+ + N_-$
subgap resistors,
\begin{equation} \label{subgap}
I_k = {n_- - n_+ \over N_+ + N_-}, \;\; -N_- < k \leq N_+,
\end{equation}
where $n_\pm(E) = n_F(E_{\pm N_\pm})$. From Eqs.\ (\ref{outgap}) and
(\ref{subgap}) we obtain the current spectral density in
Eq.\ (\ref{I2}) as $\sum_{k=-\infty}^\infty I_k = 1$, which results in
Ohm's law, $V=IR$, for the net electric current through the junction.
This conclusion is closely related to our disregarding the proximity
effect and the normal scattering at the interface. Actually, both of
these factors lead to the appearance of SGS and excess or deficit
currents in the $I$-$V$ characteristic, with the magnitude increasing
along with the interface barrier strength and the ratio
$\xi_0/d$.\cite{circuit}

The subgap populations can be found as the potentials of the nodes of
the subgap ``voltage divider'',
\begin{equation} \label{divider}
n_k =  n_- - (n_- - n_+){N_- +k\over N_+ + N_-}.
\end{equation}

By making use of Eqs.\ (\ref{S2})-(\ref{divider}), the net current
noise can be expressed through the sum of the thermal noise of
quasiparticles outside the gap,
\begin{eqnarray} \label{S>}
&\displaystyle S_> = {4T\over 3R}\Big\{ 2\left[n_F(\Delta) + n_F(\Delta
+ eV)\right] \nonumber
\\ &\displaystyle\left.+  \left[ {eV\over T} + \ln {n_F(\Delta+eV)
\over n_F(\Delta)} \right] \coth{eV \over 2T}\right\},
\end{eqnarray}
and the subgap noise,
\begin{eqnarray} \label{SD}
&\displaystyle S_\Delta = {2\over 3R}\int_0^{eV} dE (N_+ +
N_-)[f_{+-}+f_{-+} \nonumber \\ &\displaystyle\left.
+2(f_{++}+f_{--})\right], \quad f_{\alpha\beta} = n_\alpha(1-n_\beta).
\end{eqnarray}

\begin{figure}
\epsfxsize=8.0cm\epsffile{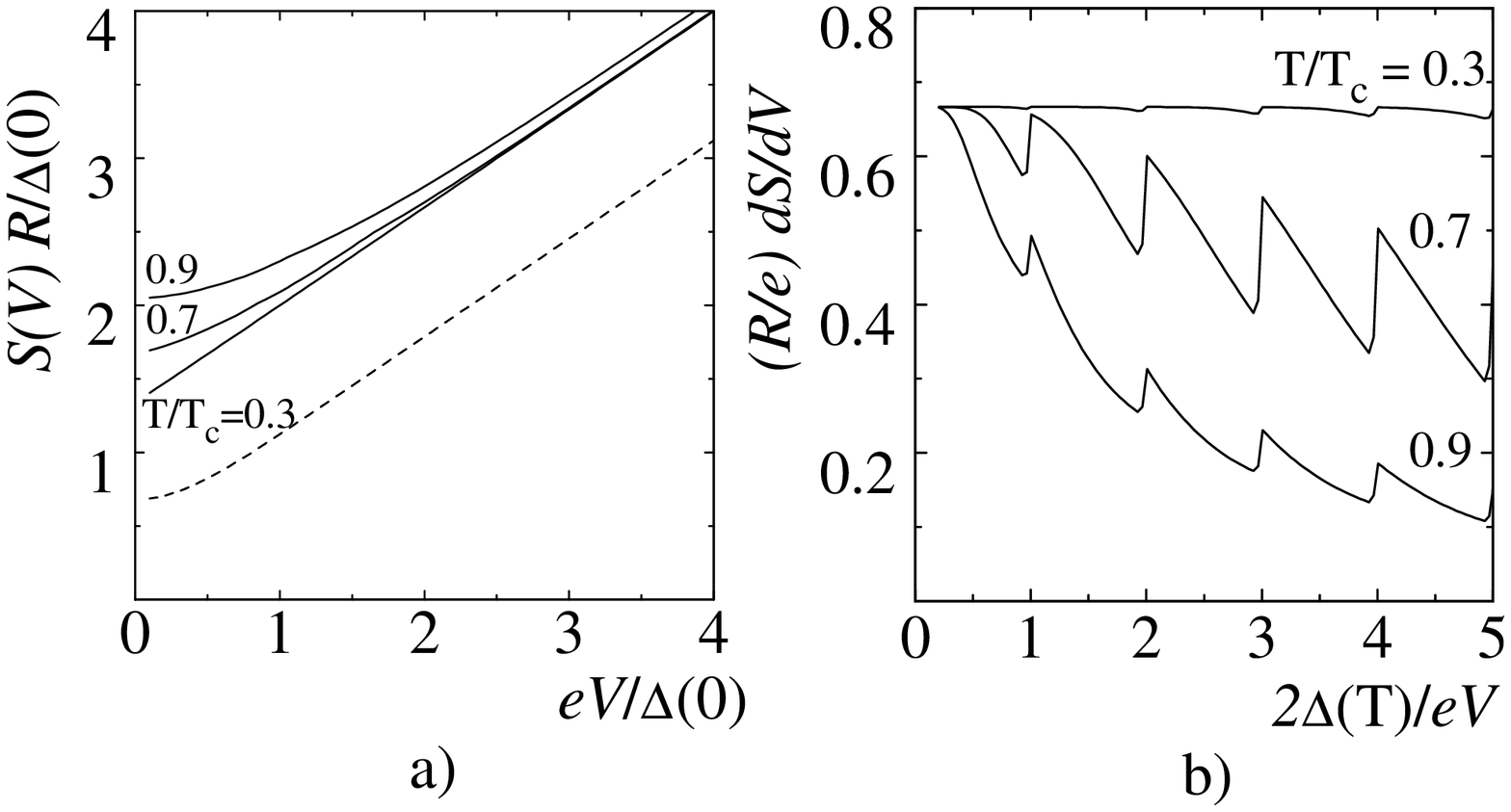} \vspace{0.2cm}
\caption{Spectral density $S$ of current noise vs voltage (a) and its
derivative $dS/dV$ vs inverse voltage (b) at different temperatures.
Dashed line shows the result for normal metal junction\cite{1/3Nagaev}
at $T=0.3T_c$.} \label{noisev}
\end{figure}

At low temperatures, $T \ll \Delta$, the thermal noise $S_>$ vanishes,
and the total noise coincides with the subgap shot noise, which takes
the form
\begin{equation} \label{T=0}
S = {2\over 3R}\int_0^{eV} dE (N_+ + N_-) = {2\over 3R}(eV + 2\Delta),
\end{equation}
of $1/3$-suppressed Poisson noise $S=(2/3)qI$ for the effective charge
$q = e(1+ 2\Delta/eV)$. At $V \rightarrow 0$, the shot noise turns to a
constant value $4\Delta/3R$ in Eq.\ (\ref{S(0)}), which is identical to
the result of Ref.\ \onlinecite{SNINS} and therefore seems to be
universal for the incoherent MAR regime in long diffusive junctions and
independent of the shot noise mechanism. At finite voltages, this
quantity plays the role of the ``excess'' noise, i.e. the
voltage-independent addition to the shot noise of a normal metal at low
temperatures [see Fig.\ \ref{noisev}(a)]. Unlike short junctions,
where the excess noise is proportional to the excess current,\cite{DC}
in our system the excess current is small and has nothing to do with
large excess noise.

Results of numerical calculation of the noise at finite temperature are
shown in Fig.\ \ref{noisev}. While the temperature increases, the noise
approaches its value for normal metal structures,\cite{1/3Nagaev}
with additional Johnson-Nyquist noise coming from thermal excitations.
In this case, the voltage-independent part of current noise may be
qualitatively approximated by the Nyquist formula $S(T) = 4 T^\ast/R$
with the effective temperature $T^\ast = T + \Delta(T)/3$. The most
remarkable phenomenon at nonzero temperature is the appearance of
steps in the voltage dependence of the derivative $dS/dV$ at the gap
subharmonics $eV = 2\Delta/n$ [Fig.\ \ref{noisev}(b)], which reflect
discrete transitions between the quasiparticle trajectories with
different numbers of Andreev reflections. The magnitude of SGS
decreases both at $T \rightarrow 0$ and $T \rightarrow T_c$, which
resembles the behavior of SGS in the $I$-$V$ characteristic of long
ballistic SNS junction with perfect interfaces within the OTBK
model.\cite{OTBK} A small ``residual'' SGS in current noise, similar to
the one in the $I$-$V$ characteristic,\cite{circuit} should occur at $T
\rightarrow 0$ due to normal scattering at the interface or due to
proximity effect [see comments to Eq.\ (\ref{subgap})].

The role of inelastic scattering is to suppress the nonequilibrium of
the subgap electrons created by MAR. Within the relaxation time
approximation,\cite{SNINS,circuit} the nonequilibrium distribution of
the subgap quasiparticles holds as soon as the inelastic
relaxation time $\tau_\varepsilon(E)$ at the
characteristic energies $E \sim \Delta$ is larger than the diffusion
time through the junction, $\tau_d(V) \sim (2\Delta/eV)^2 d^2/D$. Thus,
in this model, the enhanced shot noise (as well as the SGS in the
$I$-$V$ characteristic\cite{circuit}) can be observed if the applied
voltage is sufficiently large, $eV > 2\Delta W^{-1/2}_\varepsilon$,
where $W_\varepsilon=E_{\text{Th}} \tau_\varepsilon(\Delta)/\hbar$; at
smaller voltages, $S(V)$ should decrease and approach the thermal noise
level. The noise temperature is equal to the physical temperature $T$ if
the inelastic scattering is dominated by electron-phonon interaction
(assuming that the phonons are in equilibrium with the electron
reservoir). If the electron-electron (e-e) scattering dominates, the
noise temperature may exceed the temperature $T$ of the electron reservoir
if this temperature is small, $T\ll\Delta$ (hot electron regime). The
reason is that at low temperature, the subgap electrons are well
decoupled from the reservoir (electrons outside the gap) due to weak
energy flow through the gap edges.

In order to quantitatively analyze this situation, we consider the
small voltage limit, $eV \ll \Delta$, in equation Eq.\ (\ref{diff}),
taking into account the e-e collision term. Due to weak spatial
dependence of the population numbers at small voltage, they can be
replaced by their boundary values, $n_0^{e,h}(E) \approx n_d^{e,h}(E)
\equiv n(E)$, in the e-e collision integral $I_{ee}(n)$. This allows us
to easily  include the collision term in the recurrences of
Eq.\ (\ref{recurr}). Within the same approximation, these recurrences
are to be considered as differential relations, which results in the
diffusion equation for $n(E)$,
\begin{equation} \label{diffE}
D_E{\partial^2 n \over \partial E^2 } = I_{ee}(n).
\end{equation}
where $D_E=(eV)^2 E_{\text{Th}}/\hbar$ is the effective diffusion
coefficient in energy space. The finite resistance $R_{NS}$ of the
NS interfaces, which partially blocks quasiparticle diffusion, can be
taken into consideration by renormalization of the diffusion
coefficient, $D_E \rightarrow D_E [1+(d/\xi_0)(R_{NS}/R)^2]^{-1}$ (see
Ref.\ \onlinecite{circuit}).

Equation (\ref{diffE}) describes the crossover from the
``collisionless'' MAR regime to the hot electron regime as function of
the parameter $D_E\tau_{ee}(\Delta)/\Delta^2$.  In the hot electron
limit, $\Delta^2 \gg D_E\tau_{ee}$, the collision integral dominates in
Eq.\ (\ref{diffE}), and therefore the approximate solution of the
diffusion equation is the Fermi function with a certain effective
temperature $T_0 \ll \Delta$. The value of $T_0$ can be found from
Eq.\ (\ref{diffE}) integrated over energy within the interval
$(-\Delta,\Delta)$ with the weight $E$, taking into account the
boundary conditions $n(\pm \Delta) = n_F(\pm \Delta)$ and neglecting
the exponentially small derivative $\partial n / \partial E$ at the gap
edges. At zero temperature of the reservoir, we obtain an asymptotic
equation for $T_0$,
\begin{equation} \label{asymptT}
(eV)^2 W_\varepsilon \exp(\Delta/T_0) = T_0\Delta (1+ T_0/
\Delta),
\end{equation}
which shows that the effective temperature of the subgap electrons
decreases logarithmically with decreasing voltage. The noise of
the hot subgap electrons is given by the Nyquist formula with
temperature $T_0$,
\begin{equation} \label{asymptS}
S(V) = (4T_0/ R)\left[1 - 2 \exp(-\Delta/T_0)\right],
\end{equation}
where the last term is due to the finite energy interval available for
the hot electrons, $|E|<\Delta$. Equations (\ref{asymptT}),
(\ref{asymptS}) give a reasonably good approximation to the result of
the numerical solution of Eq.\ (\ref{diffE}).\cite{NagaevSNS}

In summary, we have calculated current noise in a long diffusive SNS
structure with low-resistive interfaces at arbitrary temperatures.
Whereas the $I$-$V$ characteristic is approximately described by Ohm's
law, the current noise reveals all characteristic features of the MAR
regime: ``giant'' enhancement at low voltages, pronounced SGS, and
excess noise at large voltages. In the limit of strong
electron-electron scattering, the junction undergoes crossover to the
hot electron regime, with the effective temperature of the subgap
electrons decreasing logarithmically with the voltage.

Support from NFR, KVA and NUTEK (Sweden), and from FRF (Ukraine) is
gratefully acknowledged.


\begin{references}
\bibitem{footnote2} The unit values of the network resistances are
provided by the normalization of ``currents'' $I_k$ (note that it
differs from the definition in Ref.\ \onlinecite{circuit} by the factor
$R^{-1}$).
\end{references}
\end{document}